\pdfoutput=1
\documentclass{bioinfo}
\copyrightyear{2015}
\pubyear{2015}
\usepackage{graphicx}
\usepackage{float}
\usepackage[skip=0pt]{subcaption}
\usepackage{caption}

\usepackage{mathtools}
\usepackage{amsmath}
\usepackage{adjustbox}
\begin{document}
\firstpage{1}
\title[Vis arXiv version]{Visualizing Regulation in Rule-based Models}
\author[Sekar \textit{et~al.}]{John A. P. Sekar\,$^{1}$, Jose-Juan Tapia\,$^{1}$ and James R. Faeder\,$^{1}$\footnote{to whom correspondence should be addressed}
}
\address{$^{1}$Department of Computational and Systems Biology, University of Pittsburgh School of Medicine, Pittsburgh, PA 15260}
\history{Received on XXXXX; revised on XXXXX; accepted on XXXXX}
\editor{Associate Editor: XXXXXXX}
\maketitle

\begin{abstract}
\section{Motivation:} 
Rule-based modeling is a powerful way to model kinetic interactions in biochemical systems. Rules enable a precise encoding of biochemical interactions at the resolution of sites within molecules, but obtaining an integrated global view from sets of rules remains challenging. Current automated approaches to rule visualization fail to address the complexity of interactions between rules, limiting either the types of rules that are allowed or the set of interactions that can be visualized simultaneously. There is a need for scalable visualization approaches that present the information encoded in rules in an intuitive and useful manner at different levels of detail.

\section{Results:} 
We have developed new automated approaches for visualizing both individual rules and complete rule-based models. We find that a more compact representation of an individual rule promotes promotes understanding the model assumptions underlying each rule. For global visualization of rule interactions, we have developed a method to synthesize a network of interactions between sites and processes from a rule-based model and then use a combination of user-defined and automated approaches to compress this network into a readable form. The resulting diagrams enable modelers 
to identify signaling motifs such as cascades, feedback loops, and feed-forward loops in complex models, as we demonstrate using several large-scale models. These capabilities are implemented within the BioNetGen framework but the approach is equally applicable to rule-based models specified in other formats.
\section{Availability:} 
The visualization tools are packaged with BioNetGen 2.2.6, which is freely available and includes source code. Documentation is available at \href{http://bionetgen.org}{http://bionetgen.org/}.
Graphs are output in the Graph Modeling Language format (GML), which is compatible with dedicated and freely-available graph layout tools such as Cytoscape (cytoscape.org) and yEd (yworks.com). 

\section{Contact:} \href{faeder@pitt.edu}{faeder@pitt.edu}
\section{Supplementary information:} 
Supplementary materials are available at \textit{Bioinformatics\/} online (including Figs.~S1-S9).
\end{abstract}

\section{Introduction}
In models of biochemical systems, a relatively small number of interactions between sites on molecules can generate combinatorially large networks of species and reactions \cite[]{Hlavacek2003}. Rule-based modeling frameworks such as BioNetGen \cite[]{Blinov2004,Faeder2009}, Kappa \cite[]{Danos2004} and Simmune \cite[]{Meier-Schellersheim2006} provide a compact specification for such reaction networks by allowing interactions to be specified at the level of sites within molecules. Rule-based modeling has increased in popularity in recent years \cite[]{Bachman2011,Chylek2014a} because these models explicitly state the site-based assumptions \cite[]{Sekar2012,Chylek2014a}, enable the automated generation and simulation of large reaction networks \cite[]{Faeder2009}, and enable network-free Monte-Carlo simulation when the implied networks are too large to generate \cite[]{Danos2007a,Yang2008,Sneddon2011}. An interchange format has been proposed recently to facilitate inter-operability between existing frameworks (SBML-multi, sbml.org) and several large libraries of signaling interactions have been constructed using BioNetGen \cite[]{Thomson2011,Sekar2012,Creamer2012,Chylek2014}. The growing number and complexity of rule-based models accentuate the need for effective visual tools that promote rapid understanding of models, model reuse, and new forms of analysis. 

\begin{figure}[t]
\centering
\subcaptionbox{Rules \label{rviz_fceri}}[0.9\linewidth]
{\includegraphics[width=0.9\linewidth,keepaspectratio]{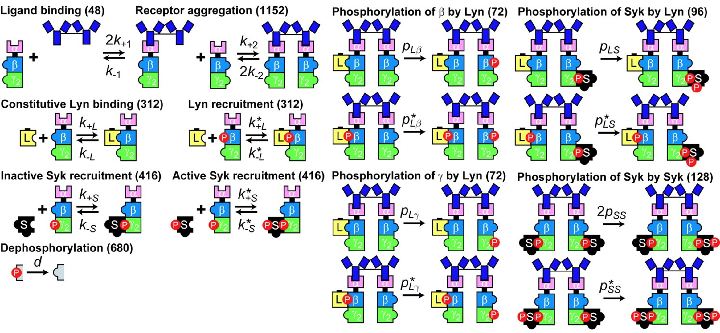}\vspace{3mm}}
\vspace{3mm}
\subcaptionbox{Reaction Network \label{rxn}}[0.45\linewidth]
{\includegraphics[width=0.45\linewidth,keepaspectratio]{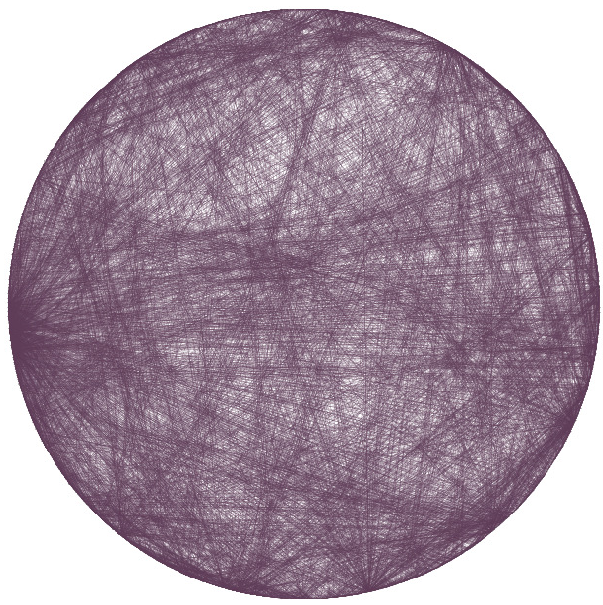}}
\subcaptionbox{Rule Influence Diagram\label{rinf}}[0.45\linewidth]
{\includegraphics[width=0.45\linewidth,keepaspectratio]{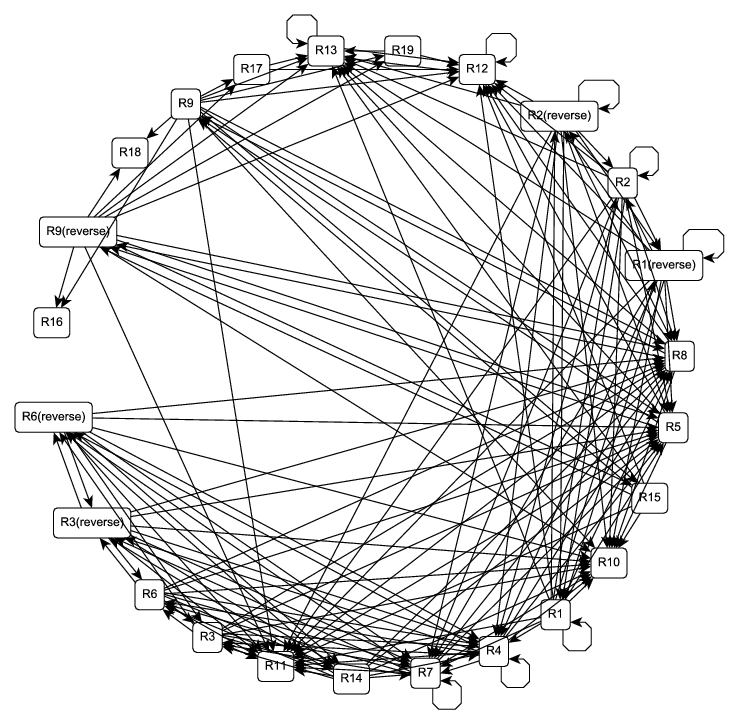}}
\subcaptionbox{Contact Map \label{cmap}}[0.45\linewidth]
{\includegraphics[width=0.45\linewidth,keepaspectratio]{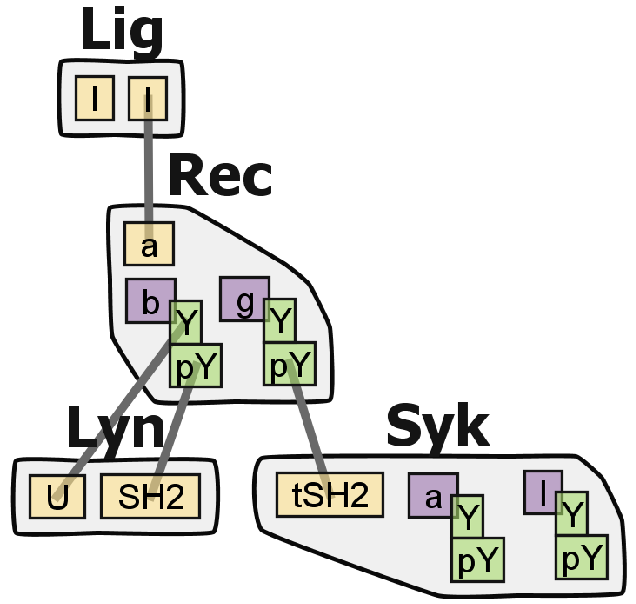}}
\subcaptionbox{Extended Contact Map\label{ecm}}[0.45\linewidth]
{\includegraphics[width=0.45\linewidth,keepaspectratio]{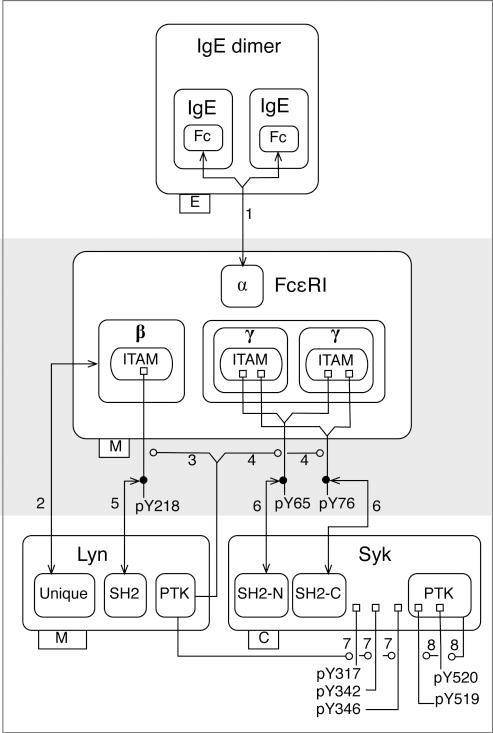}}
\caption{Previous visualizations of the \cite{Faeder2003} model. (A) Graphical representation of individual rules. (B) Reaction network with nodes representing chemical species and reactions. (C) Rule influence diagram defined by \cite{Smith2012}. (D) Contact map showing the molecules, components, states, and bonds that are present in the model. (E) Extended contact map from \cite{Chylek2011} shows contextual requirements for a subset of component state modifications, e.g. edges labeled 3, 4, 7, and 8.}
\label{fig_prior_art}
\end{figure}

The core idea underlying rule-based modeling is to use graphs to represent chemical species such as molecules and complexes and then provide the ability to specify classes of species and reactions, instead of manually specifying individual ones. This is achieved by the use of subgraph isomorphism: the \emph{pattern} graph selects a class of species that share a specified subgraph and the \emph{reaction rule} is a transformation on pattern graphs that translates to equivalent reactions on the matched species. By assigning a rate law to a reaction rule, the user simultaneously specifies the kinetics of every reaction matched by the rule, leading to a compact specification. The pattern graphs specify details at the level of individual sites on molecules, so the reaction rule is an explicitly site-based specification. This has enabled the construction of models with detailed site-based interactions, such as the model of early events in signaling through the high-affinity receptor for IgE (Fc$\varepsilon$RI), which is shown in Fig.~\ref{rviz_fceri} \cite[]{Faeder2003}. 

Although rules enable an intuitive site-based specification, conveying rules and rule-based models to a wider audience is difficult without good visualizations. Comprehending a reaction rule requires static analysis of the graphs involved: the parts of the reactant graphs that are transformed to generate the product graphs are called the \emph{reaction center}, whereas the parts preserved on both sides of the rule are called the \emph{reaction context}. Static analysis is also necessary to characterize signal flow from one rule to another. Specifically, a rule \emph{influences} a second rule if the action of the first rule modifies elements that are requirements for the second rule. For example, a phosphorylation rule would influence a binding rule if the phosphorylated state was required context for binding. Detecting influences requires sorting structures within each rule into reaction center and context and comparing pairs of rules to identify overlaps.

Visual abstractions based on reactions perform poorly in conveying the interplay of reaction center and context, both within and across rules. For example, it requires work to discern from Fig.~\ref{rviz_fceri} how the rules influence one another and which rules share a particular reaction center. Visualization of the full network of chemical species and reactions implied by the rules (Fig.~\ref{rxn}) makes the problem considerably worse. These problems apply to other visualizations that show reactants and products separately, such as the SBGN Process Description standard \cite[]{LeNovere2009} and the automated visualizations of Simmune Modeler \cite[]{Zhang2013}. Bypassing molecular structure and showing rule influences directly results in a surprisingly dense rule influence diagram (Fig.~\ref{rinf}, \cite{Smith2012}), and these influences are difficult to interpret without showing the structures involved. Other approaches to visualizing rule-based models build pathway diagrams focusing on a specific subset of influences. The Kappa story \cite[]{Danos2007,Danos2012} depicts a sequence of rules that results in a specified output, but building the story requires specific parameter choices and computationally expensive stochastic simulations. The Simmune Network Viewer \cite[]{Cheng2014} visualizes the model as a network of binding processes, but influences between rules are conveyed one-at-a-time and require user interaction.

The contact map (Fig.~\ref{cmap}) presents a compact summary of the molecular components and binding interactions present in a rule-based model \cite[]{Danos2007}, but neither the static contact map nor the interactive version \cite[]{Smith2012} presents a global view of influences between rules. The Extended Contact Map (Fig.~\ref{ecm}, \cite{Chylek2011}) and its antecedents, the SBGN Entity Relationship Diagram \cite[]{LeNovere2009} and the Molecular Interaction Map \cite[]{Kohn2006}, do superimpose elements of signal flow onto the contact map, but it rapidly becomes intractable as the number of influences increases. In addition, these diagrams must be drawn manually (requiring commercial software for optimal results) and are not directly connected to an underlying set of rules \cite[]{Chylek2011}.

Recently, \cite{Tiger2012} have introduced the \emph{regulatory graph}, which is a bipartite graph on elemental structural features (binding sites, phosphorylated states, etc.) and elemental processes (binding, phosphorylation, etc.), and which provides a global visualization for models in the {\tt rxncon} framework. The visualization is enabled by the {\tt rxncon} specification, where the modeler specifies the bipartite relationships between structures and processes in a tabular format. These are readily translated into the regulatory graph and are also processed by {\tt rxncon} to reconstruct the kinetic specification using predefined templates of rules. This approach was shown to be effective for visualizing large signaling networks \cite[]{Tiger2012}; however, the specification limits the variety of complexes and reaction rules that can be modeled explicitly. For example, neither the model of Fig.~\ref{fig_prior_art} nor the larger model libraries presented in this work can be translated into the current version of {\tt rxncon}.

In this work we present two new forms of visualization that can be applied to a general set of reaction rules such as would be specified in BioNetGen, Kappa, or Simmune. \emph{Compact rule visualization} eliminates the redundancy found in reaction-based representations of rules and makes a clear separation between reaction center and context. The \emph{rule-derived regulatory graph} provides a global visualization of rule influences mediated through elemental structural features, which are linked to rules through edges that distinguish reaction center and context relationships. Because the raw form of this graph is densely connected, we develop a set of pruning and coarse-graining procedures to improve readability. We demonstrate that rule-derived regulatory graphs can be generated using the methods described here and displayed using freely available graph layout software, even for models involving hundreds of rules. We show that the resulting diagrams can be used to identify cascades, feedback loops and feed-forward loops in the model architecture, which was not possible using previously available automatically generated diagrams.

\section{Approach}
We first present the syntax and visualization of patterns and reaction rules, which are the essential features of a rule-based model specification, and then introduce \emph{atomic patterns}, which can be used to decompose patterns into their elemental structural features. We show how this decomposition leads directly to the rule-derived regulatory graph, which can be aggregated from individual rules into a global visualization of the model, and which can be compressed with modeler input into compact pathway diagrams. The graph theory and algorithms underlying the derivation of rule and regulatory graph visualizations are presented in the Supplement and are based on definitions of the BioNetGen language from \cite{Hogg2014} and hierarchical graphs from \cite{Lemons2011}. Figure~S1 provides a visual summary for this section.

\subsection{Patterns}
\begin{figure}[t]
\centering
\subcaptionbox{Syntax \label{site_graph_syn}}[0.4\linewidth]
{
{\tt \raisebox{-3ex}{E(s!1).S(e$\sim$Y!1)}}
\begin{tabular}{ll}
& \\
{\sf \scriptsize Molecule} & {\tt E,S} \\
{\sf \scriptsize Component} & {\tt s,e} \\
{\sf \scriptsize Internal State} & {\tt $\sim$Y} \\
{\sf \scriptsize Bond} & {\tt !1} \\
\end{tabular}
}
\subcaptionbox{Site Graph \label{site_graph}}[0.5\linewidth]
{\includegraphics[width=.45\linewidth,keepaspectratio]{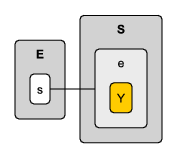}}
\label{es_patt}
\caption{Visualizing patterns. (A) Elements of the pattern syntax. The pattern shown represents an enzyme complexed with unphosphorylated substrate. (B) Visualizing as a site graph by nesting nodes and using edges to show bonds.}
\end{figure}

A \emph{pattern} in BioNetGen is a graph specifying a local arrangement of molecules, components, internal states, and bonds. The pattern selects one or more complexes that contain the specified arrangement, analogous to matching strings using regular expressions. Figure~\ref{site_graph_syn} shows a pattern that contains \emph{molecules} {\tt E} and {\tt S} representing enzyme and substrate molecules respectively, which contain substrate-binding and enzyme-binding sites respectively, represented as \emph{components} {\tt s} and {\tt e} and enclosed within brackets. The dot connecting the molecules shows that they are in the same complex. The $\sim$ symbol indicates an \emph{internal state} on a component, which is used to represent internal attributes such as covalent modifications. Here, {\tt e$\sim$Y} represents an unphosphorylated tyrosine {\tt Y} on the enzyme-binding site {\tt e}. The {\tt !} symbol indicates a bond between a pair of components and the bridged pair is identified by the bond label such as {\tt 1}. Here, the bond {\tt !1} links components {\tt s} and {\tt e}. By using different bond labels, multiple bonds can be specified. A pattern in which all components are present and all binding and internal states are explicitly defined is referred to as a \emph{species} because it selects exactly one type of complex. Figure~\ref{site_graph} shows visualization of the pattern as a \emph{site graph}, which is obtained by hierarchically nesting nodes representing molecules, components, and internal states and by showing bonds as edges between components \cite[]{Danos2012}. 

\subsection{Reaction Rules}
\begin{figure}[t]
\centering
\begin{tabular}{@{}c@{}c@{}}
	\multicolumn{2}{c}{
    	\begin{subfigure}{.9\linewidth}
        \centering
        %\fbox{E(s) + S(e$\sim$Y) -$>$ E(s!1).S(e$\sim$Y!1)}
        \caption{Syntax} 
		{\tt \raisebox{1ex}{E(s) + S(e$\sim$Y) -> E(s!1).S(e$\sim$Y!1)} }
        \label{rr_syntax}
        \end{subfigure}
    } \\
    \begin{adjustbox}{valign=t}
    	\begin{subfigure}{.4\linewidth}
        \centering
        \caption{Direct Rule Visualization}
		\includegraphics[width=\linewidth,height=2\linewidth,keepaspectratio]{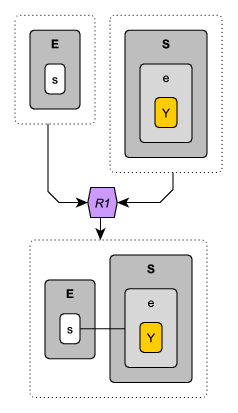}
        \label{rr_direct}
        \end{subfigure}
    \end{adjustbox} 
    & 
    \begin{adjustbox}{valign=t}
    	\begin{tabular}{@{}c@{}}
	    	\begin{subfigure}{.5\linewidth}
            \centering
        	\caption{Compact Rule Visualization}
			\includegraphics[width=.9\linewidth,keepaspectratio]{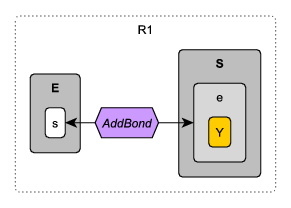}
            \label{rr_compact}
        	\end{subfigure}
            %\vspace{3mm}
            \\
            \begin{subfigure}{.5\linewidth}
            \centering
        	\caption{Rule-derived Regulatory Graph}
			\includegraphics[width=.6\linewidth,keepaspectratio]{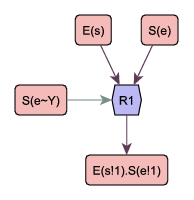}
            \label{rr_reg}
        	\end{subfigure}
        \end{tabular}
    \end{adjustbox}
    \\
\end{tabular}
\caption{Visualizing reaction rules. (A) Rule representing binding of free enzyme {\tt E(s)} and free unphosphorylated substrate {\tt S(e$\sim$Y)} to form a complex {\tt E(s!1).S(e$\sim$Y!1)}. (B) Direct visualization as a bipartite graph nesting pattern site graphs. (C) Compact visualization explicitly showing the modifications performed. (D) Rule-derived regulatory graph showing relations between classes of sites (atomic patterns) and the reaction rule. Free substrate-binding site {\tt E(s)} and enzyme-binding site {\tt S(e)} are consumed, the bond {\tt E(s!1).S(e!1)} is produced and unphosphorylated state {\tt S(e$\sim$Y)} is context. }
\label{fig_rviz}
\end{figure}

A \emph{reaction rule} in BioNetGen specifies a kinetic process whose rate law is determined by a certain arrangement of molecules, components, and bonds. In the rule, reactant patterns select the species that can participate as reactants in the process and the product patterns implicitly specify how the reacting species are transformed. The reactant patterns may match  many combinations of reacting species, so a single rule can be used to specify multiple reactions. Figure~\ref{rr_syntax} shows a reaction rule in BioNetGen syntax modeling the binding of free enzyme {\tt E(s)} and unphosphorylated substrate {\tt S(e$\sim$Y)} to form a complex {\tt E(s!1).S(e$\sim$Y!1)}.

In \emph{direct rule visualization} (Fig.~\ref{rr_direct}), the reaction rule is directly translated into a bipartite graph composed of a rule node and nodes embedding site graphs of patterns. Inflow and outflow edges on the rule node indicate reactant and product relationships respectively. This bipartite representation is standard, but has a high degree of redundancy. Structures common to both sides (reaction context) are drawn twice and can obscure the modified structures (reaction center), which hinders the rapid comprehension of rules, especially those with detailed reaction context. This motivated us to develop a more compact representation of the rule that eliminates this redundancy.

\subsection{Compact Rule Visualization}
Figure~\ref{rr_compact} presents a compact visual representation for  reaction rules that emphasizes the reaction center. The reactant and product patterns are merged together into a single site graph and graph operation nodes are added to indicate the modifications performed (see Supplement for details on the methods). Discerning reaction center from reaction context only requires locating the graph operation nodes, which are highlighted using a distinct node shape and color. In Fig.~\ref{rr_compact}, the enzyme and substrate molecules are only represented once and the AddBond graph operation node indicates that a bond is being added by the rule between components {\tt s} and {\tt e}. 
The currently supported graph operations include adding and removing bonds, adding and removing molecules and changing internal states, and examples of each operation can be found in Fig.~S2. 

\subsection{Atomic Patterns}
As mentioned previously, each pair of rules can overlap in a unique and complex manner and showing each pairwise overlap results in combinatorially complex influence diagrams such as Fig.~\ref{rinf}. \cite{Tiger2012} showed that a more fruitful approach is to represent signal flow mediated through elemental structural features. In BioNetGen, these features are types of molecules, components, internal states and bonds, which we call \emph{atomic patterns}. Each atomic pattern represents a simple biochemical observable. For example, the atomic pattern {\tt E(s!1).S(e!1)} matches all complexes containing an enzyme-substrate bond and the atomic pattern {\tt S(e$\sim$Y)} matches all substrate molecules in which the component {\tt e} is unphosphorylated (i.e. in state {\tt Y}). To represent signal flow between reaction rules and atomic patterns, we define a systematic decomposition of patterns used in reaction rules (which can be arbitrarily complex) into atomic patterns (see Supplement). For example, the patterns in the rule in Fig.~\ref{rr_syntax} are decomposed as shown in Fig.~\ref{rr_reg}. The relationship between atomic patterns and patterns is analogous to that of atoms and molecules in chemistry. For example, the formula $C_{6}H_{12}O_{6}$ represents a molecule composed of carbon, hydrogen and oxygen, but it does not tell us whether the molecule is glucose or fructose. Similarly, atomic patterns describe an elementary composition of the system being modeled, whereas patterns describe specific configurations that participate in reaction rules.

\subsection{Rule-derived Regulatory Graph}
Figure~\ref{rr_reg} presents the \emph{rule-derived regulatory graph}, which shows relationships between the reaction rule and atomic patterns. It is synthesized by decomposing the reactant and product patterns into atomic pattern instances and determining whether each instance belongs to the reaction center or reaction context (see Supplement for details). Dark colored edges indicate reaction center and light colored edges indicate reaction context relationships respectively. Within the reaction center, inflows and outflows on the rule node distinguish reactants and products. 

\begin{figure}[t]
\centering
\subcaptionbox{Regulatory Graphs of Individual Rules \label{esreg_sep}}[.9\linewidth]
{\includegraphics[width=.9\linewidth,keepaspectratio]{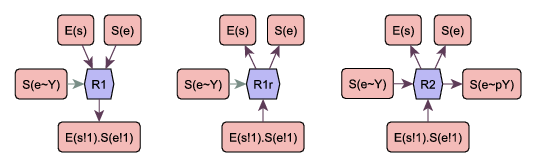}\vspace{5mm}}
\subcaptionbox{Aggregated Graph \label{esreg_all}}[.45\linewidth]
{\includegraphics[width=.45\linewidth,keepaspectratio]{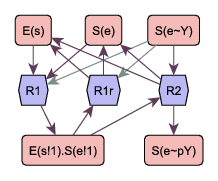}}
\subcaptionbox{Background Nodes Removed \label{esreg_nobkg}}[.45\linewidth]
{\includegraphics[width=.45\linewidth,keepaspectratio]{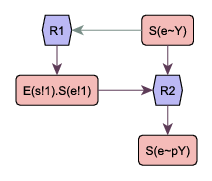}}
\caption{Aggregation and background removal in regulatory graphs. (A) Regulatory graphs of the three rules of a Michaelis-Menten mechanism. (B) Aggregating the individual graphs into a single one. (C) Removing background nodes (here, {\tt E(s)}, {\tt S(e)} and R1r) reveals the signal flow. }
\label{fig_esreg}
\end{figure}

The decomposition of reactant and product patterns into atomic patterns makes the rule-derived regulatory graph a systematic approximation of the explicitly specified reaction rule. However, since atomic patterns overlap in an all-or-none fashion, this enables a simple bipartite graph representation over the full set of rules and atomic patterns, as we show in Fig.~\ref{fig_esreg}. Figure~\ref{esreg_sep} shows rule-derived regulatory graphs of individual rules of the Michaelis-Menten mechanism, with reactant and product patterns decomposed into atomic patterns. Figure~\ref{esreg_all} is an aggregated graph constructed by merging the individual graphs and represents the complete set of interactions. Although the number of edges on this graph scales linearly with the number of rules, we still found it to be more complex than diagrams hand-drawn by experts. This motivated us to develop graph reductions to improve readability.

\subsection{Background Removal}
We observed that a number of nodes on the rule-derived regulatory graph  add visual clutter but do not provide additional insight. For example, a bond necessarily implies the existence of the binding sites that compose the bond and an unbinding rule is often the exact reverse of a binding rule. Additionally, some rules model background processes that are necessary for the kinetic specification, but do not contribute to the intuition of signal flow. Removing these nodes can improve visual comprehension, so we provide an algorithm that identifies and removes background nodes by making certain principled assumptions. Because we anticipate that these assumptions will not work for all models, we provide in our implementation the option to modify the algorithm's choices based on user input (see Documentation). Removing the free binding sites and the unbinding rule from Fig.~\ref{esreg_all} results in Fig.~\ref{esreg_nobkg}, which is more amenable to visual analysis. For example, we can identify that binding (R1) produces {\tt E(s!1).S(e!1)}, which is required for phosphorylation (R2), resulting in a net positive influence from R1 to R2. On the other hand, R2 consumes the unphosphorylated state {\tt S(e$\sim$Y)}, which is required as context for R1, resulting in a net negative influence from R2 to R1. We use background removal throughout the remainder of this work to facilitate visual analysis.

\subsection{Grouping and Collapsing}
\begin{figure}[t]
\centering
\subcaptionbox{No Grouping \label{abx1}}[.9\linewidth]
{\includegraphics[width=.9\linewidth,keepaspectratio]{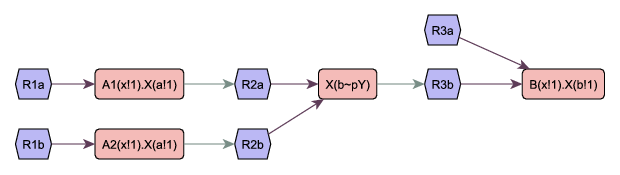}}
\subcaptionbox{Automated Rule Grouping \label{abx2}}[.9\linewidth]
{\includegraphics[width=.9\linewidth,keepaspectratio]{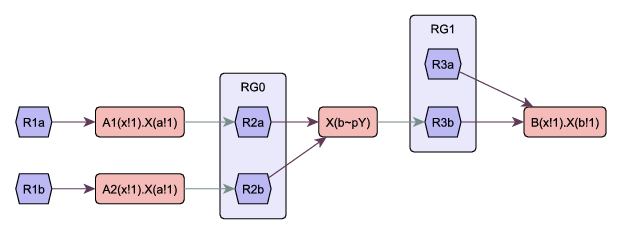}}
\subcaptionbox{Automated Rule Grouping with Expert Input \label{abx3}}[.9\linewidth]
{\includegraphics[width=.9\linewidth,keepaspectratio]{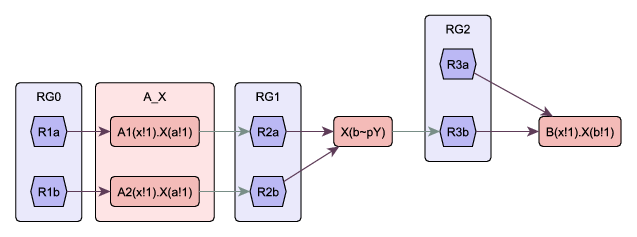}}
\subcaptionbox{Collapsing Group Nodes \label{abx4}}[.9\linewidth]
{\includegraphics[width=.9\linewidth,keepaspectratio]{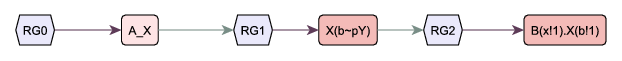}}
\caption{Grouping and collapsing on the regulatory graph. (A) Regulatory graph of a model, where X is a scaffold, A1, A2 are kinases that modulate the binding site for B on X. (B) Automated rule grouping identifies groups of rules with identical reaction centers. (C) Groups of atomic patterns can be provided as expert input. This information will be used by the algorithm when assigning rule groups. (D) Collapsing groups of nodes to single nodes reduces the size of the graph and coarse-grains the contextual interactions. }
\label{fig_abx}
\end{figure}

A typical model may contain multiple contextual variants of the same kinetic process, for example, the same pair of molecules might bind at different rates when present in different conformational states. These variants can be identified as having identical reaction centers but varied reaction contexts, and we provide an algorithm to automatically identify such groups on the regulatory graph. Figure~\ref{abx1} shows a regulatory graph of a model with four proteins A1, A2, B and X. A1 and A2 are kinases which bind X and phosphorylate a site on X that binds B. In Fig.~\ref{abx2}, the \emph{automated rule grouping} algorithm identifies that rules R2a/R2b both produce {\tt X(b$\sim$pY)} and rules R3a/R3b both produce {\tt B(x!1).X(b!1)}. A1 and A2 are two different molecule types, so the grouping algorithm does not recognize their similarity. However, this information can be provided as \emph{expert input}. For example, in Fig.~\ref{abx3}, specifying {\tt A1(x!1).X(a!1)} and {\tt A2(x!1).X(a!1)} as a group labeled A\_X results in the algorithm automatically grouping rules R1a/R1b that produce A\_X.

Groups on the regulatory graph represent higher order classes of atomic patterns and rules and these are often more important for comprehension than individual atomic patterns and rules. A graph involving these higher order classes can be achieved by \emph{collapsing group nodes}, as in Fig.~\ref{abx4}. This involves combining edges on individual group members, remapping them to a generic node representing the group and then removing individual group members. While collapsing results in a drastic reduction in graph size and complexity, it also significantly coarse-grains the context relations. A context edge on a collapsed group node implies that at least one member of the group on the uncollapsed graph has a similar edge, but it does not specifically indicate which members of the group do.

\section{Results}
\subsection{Visualizing Interactions of Reaction Rules}
\begin{figure}[t]
\centering
\subcaptionbox{Lyn-Rec binding \label{4rules1}}[.9\linewidth]
{\includegraphics[width=.9\linewidth,keepaspectratio]{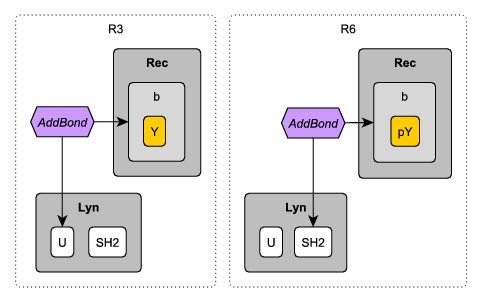}\vspace{3mm}}
\subcaptionbox{Rec phosphorylation \label{4rules2}}[.9\linewidth]
{\includegraphics[width=.9\linewidth,keepaspectratio]{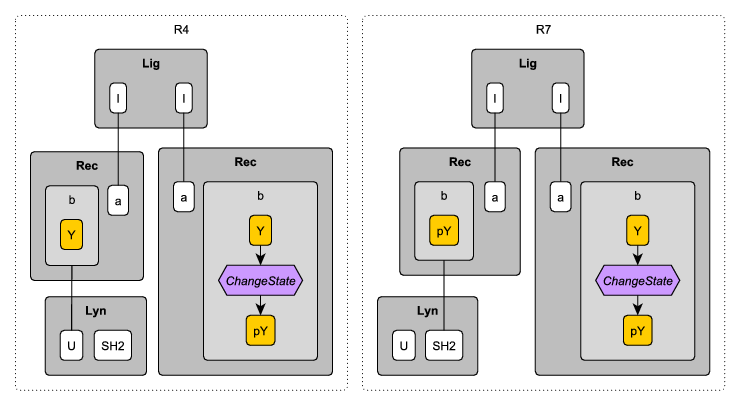}\vspace{3mm}}
\subcaptionbox{Regulatory Graph \label{4rules3}}[.9\linewidth]
{\includegraphics[width=.9\linewidth,keepaspectratio]{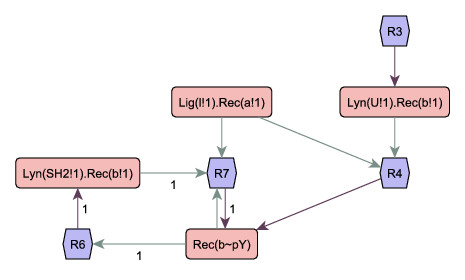}}
\caption{Rules modeling interaction of Fc$\varepsilon$RI receptor and Lyn. (A) Constitutive and activated recruitment of Lyn to receptor. (B) Phosphorylation of receptor by Lyn recruited to cross-linked dimer. (C) Regulatory graph showing the positive feedback loop between activated recruitment and receptor phosphorylation. }
\label{fig_fceri_ji_regs}
\end{figure}

Rule visualization promotes understanding the structural and kinetic assumptions encoded in the model. To demonstrate this we use four rules from \cite{Faeder2003} modeling the interaction of cytoplasmic Lyn kinase with the Fc$\varepsilon$RI receptor. Figure~\ref{4rules1} shows two rules R3 and R6 modeling binding of Lyn to receptor. R3 models constitutive recruitment: the binding site on the receptor is unphosphorylated and the binding domain on Lyn is {\tt Lyn(U)}. R6 models activated recruitment: the binding site on the receptor is phosphorylated and the binding domain on Lyn is {\tt Lyn(SH2)}. Figure~\ref{4rules2} shows two rules R4 and R7 modeling trans-phosphorylation of receptor by Lyn recruited to the cross-linked dimer. In R4, the Lyn kinase is constitutively recruited and in R7, it is actively recruited. 

The structures common to each pair of rules mediate the interactions between them. R4 requires constitutive binding, which is produced by R3. R6 requires phosphorylation, which is produced by both R4 and R7. R7 requires activated binding, which is produced by R6. Figure~\ref{4rules3} shows the rule-derived regulatory graph that summarizes these relations. It also enables identifying the positive feedback loop in the system between receptor phosphorylation and Lyn recruitment (edges labeled 1 in Fig.~\ref{4rules3}). Such network-level motifs, which are important for conveying the architecture and function of the modeled system, can be rapidly identified on the regulatory graph, while they are typically hidden or obscured in other methods for visualizing rule-based models.

\subsection{Organizing and Visualizing Pathways}
\begin{figure}[t]
\centering
\subcaptionbox{ \label{fceri_ji_reg1}}[.9\linewidth]
{\includegraphics[height=.8\linewidth,width=.9\linewidth,keepaspectratio]{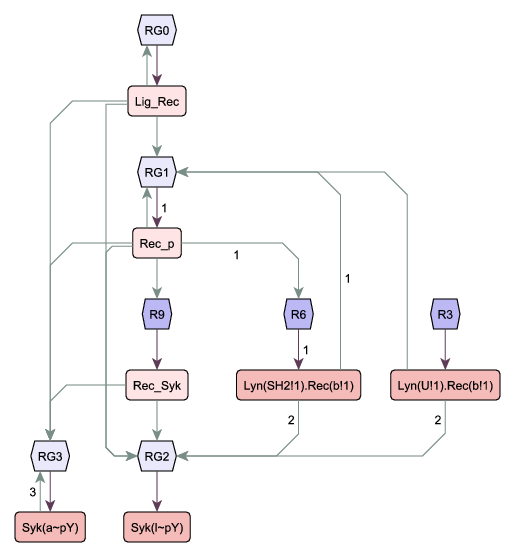}}
\subcaptionbox{ \label{fceri_ji_reg2}}[.5\linewidth]
{\includegraphics[height=.8\linewidth,width=.5\linewidth,keepaspectratio]{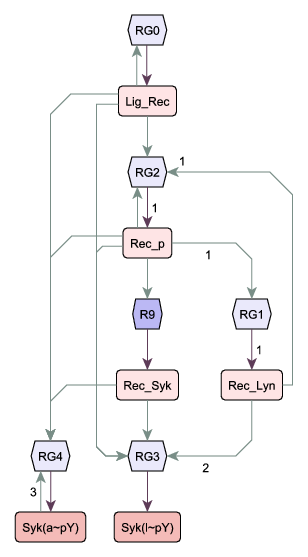}}
\subcaptionbox{ \label{fceri_ji_reg3}}[.4\linewidth]
{\includegraphics[height=.8\linewidth,width=.4\linewidth,keepaspectratio]{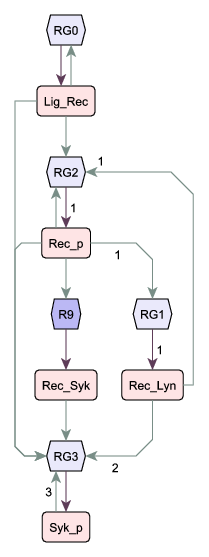}}
\caption{Successive coarse-grainings of the regulatory graph of the \cite{Faeder2003} model. (A) Both Lyn-receptor binding states and both Syk phosphorylation sites are resolved. (B) The Lyn-receptor binding states were merged under Rec\_Lyn. (C) The Syk phosphorylation sites were merged under Syk\_p. }
\label{fig_4rules}
\end{figure}

The regulatory graph can be organized using expert input to produce compact pathway diagrams of complex models. To demonstrate this, we use the \cite{Faeder2003} model, whose prior visualizations are shown in Fig.~\ref{fig_prior_art}. From the contact map in Fig.~\ref{cmap}, we see that the model has a receptor, a ligand that binds receptor, and two cytoplasmic kinases Lyn and Syk that are recruited to the $\beta$ and $\gamma$ sites on the receptor respectively. The full rule-derived regulatory graph, inferred without any user intervention, is presented in the Fig.~S3. The graph in Fig.~\ref{fceri_ji_reg1} was generated from the full graph by removing background, providing a particular grouping strategy, and collapsing group nodes. Specifically, the ligand-receptor bond was placed under the label Lig\_Rec, the receptor-Syk bond was placed under Rec\_Syk, and the two receptor phospho-sites were grouped under Rec\_p. In Fig.~\ref{fceri_ji_reg1}, we see that the algorithm automatically grouped ligand-binding rules under RG0 and receptor phosphorylation rules under RG1. 

The grouping strategy affects the resolution with which the regulatory architecture is presented. First, we demonstrate that leaving atomic patterns ungrouped allows us to distinguish regulatory features specific to each atomic pattern. In Fig.~\ref{fceri_ji_reg1}, we did not group the two Lyn-receptor binding states or the two Syk phospho-sites. Consequently, we are able to discern constitutive and activated Lyn binding processes (R3 and R6 respectively) on the regulatory graph, and we can see that activated binding alone participates in a feedback loop with receptor phosphorylation (edges labeled 1). Similarly, we can identify Syk phosphorylation specific to each phospho-site (RG2 and RG3 in Fig.~\ref{fceri_ji_reg1}) and highlight the contextual differences between them: one of them is Lyn-mediated (edges labeled 2) and the other is Syk-mediated (edge labeled 3).

Next, we demonstrate how we can achieve a controlled loss of resolution by grouping and collapsing. Figures~\ref{fceri_ji_reg2}~and~\ref{fceri_ji_reg3} show successive coarse-grainings on the map in Fig.~\ref{fceri_ji_reg1}. In Fig.~\ref{fceri_ji_reg2}, we grouped the two atomic patterns modeling Lyn-receptor binding under Rec\_Lyn. This renders them indistinguishable on the collapsed graph, resulting in a single Lyn-binding group (RG1 in Fig.~\ref{fceri_ji_reg2}) through which the feedback loop is now routed. In Fig.~\ref{fceri_ji_reg3}, we additionally grouped the two Syk phospho-sites under Syk\_p. This results in a single Syk phosphorylation group (RG3 in Fig.~\ref{fceri_ji_reg3}) and no differentiation between Lyn-mediated and Syk-mediated phosphorylation. The uncollapsed versions of the graphs in Figs.~ \ref{fceri_ji_reg1}-\ref{fceri_ji_reg3} are presented in Fig.~S4. 

The grouping on atomic patterns, followed by the automated grouping of rules mimics the organization of biochemical knowledge that an expert would perform during the manual diagramming process. Since the grouping is applied formally to the rule-derived regulatory graph that is derived automatically from the model, the correspondence between the model and the visualization is always preserved. A useful grouping strategy is one that accounts for the nuances of the specific model, the purpose of the diagram, and the level of detail that is appropriate for the intended audience. For example, Figs.~\ref{fceri_ji_reg2}~or~\ref{fceri_ji_reg3} are adequate for showing an accessible pathway diagram, but Fig.~\ref{fceri_ji_reg1} is necessary to discuss Lyn and Syk mechanisms in detail. The naming of groups can also be used to improve accessibility, e.g. by using shorthand such as Lig\_Rec and Rec\_p.

\subsection{Visualizing Extended Networks}
\begin{figure}[t]
\centering
\includegraphics[width=.9\linewidth,keepaspectratio]{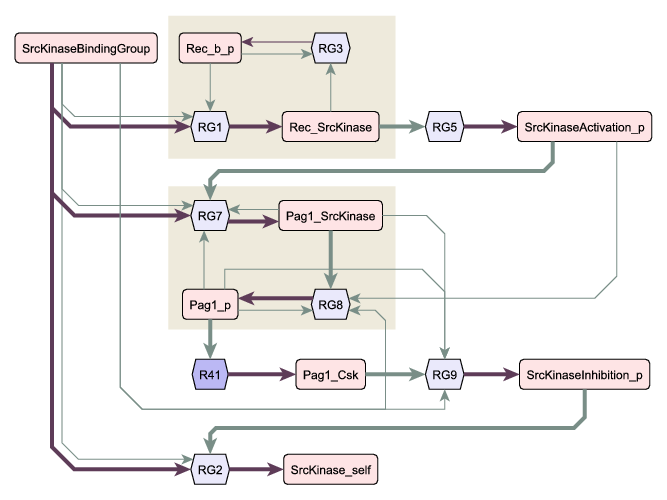}
\caption{Subset of the \cite{Chylek2014} model of Fc$\varepsilon$RI signaling showing interactions between the receptor, the Src kinases Lyn and Fyn, the scaffold Pag1 and the kinase Csk. The thick edges follow the canonical flow of signal through the network. The boxes mark positive feedback loops.}
\label{srckinase}
\end{figure}

The process of inferring, strategically grouping and reducing the regulatory graph is scalable to large rule-based models. Here, we consider the Fc$\varepsilon$RI receptor signaling library constructed by \cite{Chylek2014} and the signaling model of the ErbB receptor family constructed by \cite{Creamer2012}. The models have 17 and 19 molecule types respectively (Fig.~S5) and 178 and 625 reaction rules respectively. We have generated regulatory graphs for both models by providing appropriate expert input and grouping and collapsing (Figs.~S6-S7). Figure \ref{srckinase} shows a subset of the regulatory graph of the \cite{Chylek2014} model involving the Fc$\varepsilon$RI receptor, the Pag1 scaffold and the Lyn, Fyn and Csk kinases. 

Attempting to visualize these large models provides insight into designing an appropriate grouping strategy given expert knowledge about the system. In general, grouping is guided by structural similarity, for example, one would group multiple binding modes between the same pair of molecules or multiple phosphorylation sites on the same molecule. However, as we demonstrate in Fig.~\ref{srckinase}, grouping functionally similar structures across molecules can also be useful. Lyn and Fyn have similar structure and function, so we group homologous sites on Lyn and Fyn under the common heading of SrcKinase. Binding domains on the SrcKinase work in concert to either bind a target such as receptor or scaffold, or to bind each other and exist in a self-inhibited state. These domains are grouped under the label SrcKinaseBindingGroup. Phosphorylation sites on the SrcKinase are functionally classified as activation-related and inhibition-related and grouped accordingly.

Annotating or highlighting nodes, edges, cascades and loops in the system can help elucidate a complex network of interactions. In Fig.~\ref{srckinase} the thick edges mark the canonical flow of signal through the network. SrcKinase binding to receptor (RG1) results in activation-related phosphorylation of the SrcKinase (RG5). This is an important branch point for signaling in this system because the active SrcKinase is implicated in activation of other pathways (not shown). What follows next is an inactivation cascade: SrcKinase binds to Pag1 (RG7), leading to Csk-dependent phosphorylation of inhibition-related sites (RG9) and eventually auto-inhibition of the SrcKinase (RG2). The boxes were added manually to highlight positive feedback loops in the flow that enhance binding of the SrcKinase to receptor (RG1) or Pag1 (RG7). 

After using the diagram to explain the system architecture, a discussion of the dynamical effects can then follow. For example, in the context of BCR signaling. which uses the same architecture, \cite{Barua2012} suggest that the delayed initiation of the dominant inactivation cascade results in pulse-like signaling from the Src kinases, with rapid onset and shutoff. \cite{Barua2012} also find that that for some initial concentrations of the Src kinases, this architecture can give rise to oscillations. In Fig.~S8, we present two other subsets of the \cite{Chylek2014} graph that showcase coherent and incoherent feed-forward loops in the network architecture.

\subsection{Readability Analysis}
By showing interactions between rules and structures, the regulatory graph conveys more information than the structure-centric contact map and the rule-centric rule influence diagram. However, we wanted to investigate if the regulatory graph was empirically more readable. Following \cite{Ghoniem2005}, the readability of node-link diagrams is known to decay with graph size and edge density. We compared contact maps, rule influence diagrams, and the full regulatory graphs (no background removal, no grouping) of ten published rule-based models with model sizes ranging from 24-625 rules. As expected, we found that the contact map was always the smallest representation. The regulatory graph is marginally larger than the rule influence diagram (1-2x nodes), but is much less cluttered (0.1-0.7x edges per node), suggesting that the full regulatory graph is usually more readable than the rule influence diagram and at least as readable in the worst case. We also evaluated the reduced regulatory graphs we generated for this paper for the models of \cite{Faeder2003}, \cite{Creamer2012}, and \cite{Chylek2014}. We find that the reduced regulatory graph outperforms all fully automated approaches and is even smaller than the contact map (0.25-0.6x nodes). We present this data in Fig.~S9.

\section{Discussion}
In this work, we have developed visual tools for rule-based models at different levels of detail, from individual rules to the whole model. We have adopted a strategy of systematic coarse-graining that enables global visualization: explicitly specified rules are simplified to rule-derived regulatory graphs that can be aggregated and further reduced with expert input. The resulting pathway diagrams enable the identification of network features such as cascades, feedback loops and feed-forward loops, which are critical for understanding model architecture and function. We provide an implementation in the BioNetGen software package, but the underlying theory is applicable to related rule-based frameworks such as Kappa \cite[]{Danos2004} and Simmune \cite[]{Meier-Schellersheim2006}.

In ongoing efforts to standardize the encoding of all biochemical knowledge \cite[]{Demir2010,Cohen2015}, rule-based languages are playing an important role by enabling the explicit encoding of site-based hypotheses of biochemical interactions. As we show here, the rule-derived regulatory graph provides an effective visual representation of model libraries with hundreds of interactions. We expect that there will eventually be central databases of rules targeting whole cells and organisms and that global visualizations such as the rule-derived regulatory graph will be necessary for effective maintenance and usage of such libraries. There are several areas for future development, such as using higher-order patterns (i.e. non-atomic), supporting transport operations, incorporating patterns used in rate laws \cite[]{Sneddon2011}, using more complex grouping strategies \cite[]{Vehlow2015}, developing alternative views of the graph using matrices \cite[]{Tiger2012} or hierarchies \cite[]{Hu2013}, superimposing dynamical quantities like reaction fluxes \cite[]{Konig2010}, and the derivation of Boolean models from the regulatory graph \cite[]{Mori2015}. We are also currently working on adapting the regulatory graph to existing standards for representing pathways \cite[]{Demir2010} and visualizations \cite[]{VanIersel2012}.

\section*{Acknowledgement}
We thank Justin S. Hogg, Leonard A. Harris and William S. Hlavacek for helpful discussions and comments and Robert P. Sheehan and Cihan Kaya for testing the visual tools.

\paragraph{Funding\textcolon} 
NIH grant P41 GM103712 and NSF Expeditions in Computing Grant (award 0926181).

\bibliographystyle{natbib}
\bibliography{vizrefs3}

\end{document}